\documentstyle[epsfig]{elsart}
\begin{document}
\runauthor{STAR Collaboration }
\begin{frontmatter}
\title{Identified hadron spectra at large transverse momentum in $p$+$p$ and
$d$+Au collisions at $\sqrt{s_{\mathrm {NN}}}$ = 200 GeV}

%%%%% Authorlist taken from STAR Web page %%%%%%
 \author[uuk]{J.~Adams},
 \author[pu]{M.M.~Aggarwal},
 \author[vecc]{Z.~Ahammed},
 \author[kent]{J.~Amonett},
 \author[kent]{B.D.~Anderson},
 \author[ucd]{M.~Anderson},
 \author[dubna]{D.~Arkhipkin},
 \author[jinr]{G.S.~Averichev},
 \author[jammu]{S.K.~Badyal},
 \author[nikhef]{Y.~Bai},
% ...

(STAR Collaboration)

%%%%% Institute list taken from STAR Web page %%%%%%
% ...
\address[uuk]{University of Birmingham, Birmingham, United Kingdom}
% ...
\address[ucd]{University of California, Davis, California 95616}
% ...
\address[jinr]{Laboratory for High Energy (JINR), Dubna, Russia}
\address[dubna]{Particle Physics Laboratory (JINR), Dubna, Russia}
% ...
\address[jammu]{University of Jammu, Jammu 180001, India}
\address[kent]{Kent State University, Kent, Ohio 44242}
% ...
\address[nikhef]{NIKHEF and Utrecht University, Amsterdam, The Netherlands}
% ...
\address[pu]{Panjab University, Chandigarh 160014, India}
% ...
\address[vecc]{Variable Energy Cyclotron Centre, Kolkata 700064, India}
% ...

%\author{STAR Collaboration}
%\address{STAR Collaboration}

\date{\today}
\begin{abstract}
We present the transverse momentum ($p_{\mathrm T}$) spectra for
identified charged pions, protons and anti-protons from $p$+$p$ and
$d$+Au collisions at $\sqrt{s_{\mathrm {NN}}}$ = 200 GeV.  The spectra
are measured around midrapidity ($\mid$$y$$\mid$ $<$ 0.5) over the range
of 0.3 $<$ $p_{\mathrm T}$ $<$ 10 GeV/$c$ with particle identification
from the ionization energy loss and its relativistic rise in the Time
Projection Chamber and Time-of-Flight in STAR. The charged pion and
proton+anti-proton spectra at high $p_{\mathrm T}$ in $p$+$p$ and
$d$+Au collisions are in good agreement with a phenomenological model
(EPOS) and with next-to-leading order perturbative quantum
chromodynamic (NLO pQCD) calculations with a specific fragmentation
scheme and factorization scale. We found that all proton, anti-proton
and charged pion spectra in $p$+$p$ collisions follow
$x_{\mathrm T}$-scaling for the momentum range where particle
production is dominated by hard processes
($p_{\mathrm T}$ ${}^>_{\sim}$ 2 GeV/$c$).
The nuclear modification factor around midrapidity is found to
be greater than unity for charged pions and to be even larger for
protons at 2 $<$ $p_{\mathrm T}$ $<$ 5 GeV/$c$.
\end{abstract}
\end{frontmatter}

\section{Introduction}

%%%%%%%%%%%%%%%%%%%%%%%%%%%%%%%%%%%%%%%%%%%%%%%%%%%%%%%%%%%%%%%%
%What we can learn in p+p collisions and d+Au collisions and pQCD
%%%%%%%%%%%%%%%%%%%%%%%%%%%%%%%%%%%%%%%%%%%%%%%%%%%%%%%%%%%%%%%%%%
%%%%%%%%%%%%%%%%%%%%%%%%%%%
%Fragmentation function
%%%%%%%%%%%%%%%%%%%%%%%%%%%

The study of identified hadron spectra at large transverse momentum
($p_{\mathrm T}$) in $p$+$p$ collisions can be used to test the
predictions from perturbative quantum chromodynamics
(pQCD)~\cite{pqcd}. In the framework of models based on QCD, the
inclusive production of single hadrons is described by the
convolution of parton distribution functions (PDFs), parton
interaction cross-sections and fragmentation functions (FFs).
The PDF provide the probability of finding a parton (a
quark or a gluon) in a hadron as a function of the fraction of the
hadron's momentum carried by the parton. The FFs~\cite{ff} give the
probability for a hard scattered parton to fragment into a hadron of
a given momentum fraction. These are not yet calculable from the
first principles and hence are generally obtained from experimental
data (e.g., $e^{+}$+$e^{-}$ collisions). The factorization theorem
for cross-sections assumes that FFs are independent of the process
in which they have been determined and hence represent a universal
property of hadronization. It is therefore possible to make
quantitative predictions for other types of collision systems (e.g.,
$p$+$p$ ). Comparisons between experimental data and theory can help
to constrain the quark and gluon FFs that are critical to
predictions of hadron spectra in $p$+$p$, $p$+A, and A+A collisions.
The simultaneous study of identified hadron $p_{\mathrm T}$ spectra
in $p$+$p$ and $d$+Au collisions may also provide important
information on the PDFs~\cite{pdf} of the nucleus.

%%%%%%%%%%%%%%%%%%%%%%%%%%%%%%%%%%%%%%%%%%%%%%%%%%%%%%%%%%
%How is our study important with respect to AA collisions
%%%%%%%%%%%%%%%%%%%%%%%%%%%%%%%%%%%%%%%%%%%%%%%%%%%%%%%%%
The identified particle spectra in $p$+$p$ and $d$+Au collisions
also provide reference spectra for particle production at high
$p_{\mathrm T}$ in Au+Au collisions. Moreover, studies of identified
particle production and their ratios as a function of
$p_{\mathrm T}$ in high-energy heavy-ion collisions have revealed
many unique features in different $p_{\mathrm T}$
regions~\cite{rhicwhitepapers,hydro,reco,star_phenix_sup} and
between baryons and mesons~\cite{baryon_meson}. A good description
of both identified pion and proton spectra in $p$+$p$ and $d$+Au
collisions at intermediate and high $p_{\mathrm T}$ by NLO pQCD will
provide a solid ground for models based on jet
quenching~\cite{jquench} and quark recombination~\cite{reco}. These
emphasize the need for a systematic study of $p_{\mathrm T}$ spectra
from $p$+$p$ and $d$+Au collisions at the same energy as the
nucleus-nucleus collisions.

%%%%%%%%%%%%%%%%%%%%%%%%%%%%%%%%%%%
% What is presented in this paper
%%%%%%%%%%%%%%%%%%%%%%%%%%%%%%%%%%%
In this letter, we present the $p_{\mathrm T}$ spectra for
identified pions, protons and anti-protons in $p$+$p$ and $d$+Au
collisions at $\sqrt{s_{\mathrm {NN}}}$ = 200 GeV as measured
by the STAR experiment at RHIC. The results are compared to NLO
pQCD calculations and a phenomenological model. We also study
the $x_{\mathrm T}$-scaling in $p$+$p$ collisions
and the nuclear modification factors in $d$+Au collisions.

\section{Experiment and Analysis}
%%%%%%%%%%%%%%%%%%%%%%%%%%%%%%%
% General detector description
%%%%%%%%%%%%%%%%%%%%%%%%%%%%%%
The STAR experiment consists of several detectors to measure hadronic
and electromagnetic observables spanning a large region of the available
phase space at RHIC. The detectors used in the present analysis are the
Time Projection Chamber (TPC), the Time-Of-Flight (TOF) detector, a set of
trigger detectors used for obtaining the minimum bias data, and the
Forward Time Projection Chamber for the collision centrality
determination in $d$+Au collisions.
The details of the design and other characteristics of the detectors
can be found in Ref.~\cite{starnim}.

%%%%%%%%%%%%%%%%%%%%%%%%%%%%%%%%%%%%%%%%%
%Analysis details in brief with references
%%%%%%%%%%%%%%%%%%%%%%%%%%%%%%%%%%%%%%%%%%

A total of 8.2 million minimum bias $p$+$p$ collision events and 11.7 million
$d$+Au collision events have been analyzed for the present study.
The data set was collected during the years 2001 and 2003.
The details of minimum bias trigger conditions for $p$+$p$ and $d$+Au
collisions can be found in the Refs.~\cite{star_rdau,star_pid}.
The minimum-bias trigger captured $95\pm3\%$ of the $2.21\pm0.09$ barn $d$+Au
inelastic cross-section. The trigger efficiency was determined from a cross
study of two sets of trigger detectors: two Zero-Degree Calorimeters (ZDCs)
and two beam-beam counters (BBCs).
The absolute cross-section is derived from a Monte Carlo Glauber calculation.
These results are consistent with other recent measurements~\cite{swhite}.
The trigger for the minimum bias $p$+$p$ collisions required a coincidence
measurement of the two BBCs covering 3.3 $<$ $\mid\eta\mid$ $<$ 5.0~\cite{bbc}.
This trigger was sensitive to color exchange hadronic and doubly-diffractive
events; here, these are labelled "non-singly diffractive (NSD) events".
Using PYTHIA(v6.205)~\cite{pythia} and HERWIG~\cite{herwig}, it was determined
that the trigger measured 87\% of the $30.0\pm 3.5$ mb NSD cross-section,
which was measured via a vernier scan~\cite{vander}.
%%%%%%%%%%%%%% Fig. 1 %%%%%%%%%%%%%%%%%%%%%%%%%%%%%
\begin{figure}
\begin{center}
\includegraphics[scale=0.4]{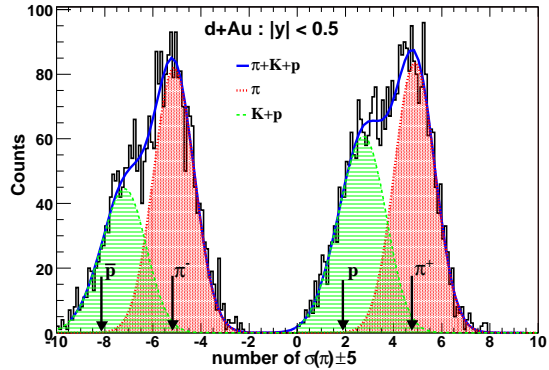}
%\vspace{-1.0 cm}
\caption{$dE/dx$ distribution normalized by pion $dE/dx$ at
4.5 $<$ $p_{\mathrm T}$ $<$ 5.0 GeV/$c$ and $\mid\eta\mid$ $<$ 0.5,
and shifted by $\pm$5 for positively and negatively charged particles,
respectively. The distributions are for minimum bias $d$+Au collisions.
The pion, proton and anti-proton peak positions are indicated by arrows.
}
\label{fig0}
\end{center}
\end{figure}
%%%%%%%%%% End of Fig.1 %%%%%%%%%%%%%%%%%%%%%%%%%%%%%
%%%%%%%%%%%%%%%%%%%%%%%%%%%%
%Particle identification
%%%%%%%%%%%%%%%%%%%%%%%%%%%%%
The data from TOF are used to obtain the identified hadron spectra
for $p_{\mathrm T}<2.5$ GeV/$c$. The procedure for particle
identification in TOF has been described in Ref.~\cite{star_tof}.
For $p_{\mathrm T}>2.5$ GeV/$c$, we use data from the TPC.
Particle identification at high $p_{\mathrm T}$ in the TPC
comes from the relativistic rise of the ionization energy
loss ($dE/dx$). Details of the method are described in Ref.~\cite{rdEdx}.
At $p_{\mathrm T}{}^{>}_{\sim} 3$ GeV/$c$, the pion $dE/dx$ is
about 10--20\% higher than that of kaons and protons due to the
relativistic rise, resulting in a few standard deviations (1-3$\sigma$)
separation between them. Since pions are the dominant component of the
hadrons in $p$+$p$ and $d$+Au collisions at RHIC, the prominent pion peak
in the $dE/dx$ distribution is fit with a Gaussian to extract the pion
yield~\cite{rdEdx}. The proton yield is obtained by integrating
the entries ($Y$) in the low part of the $dE/dx$ distribution about $2.5\sigma$
away from the pion $dE/dx$ peak. The integration limits were varied to
check the stability of the results. Fig.~\ref{fig0} shows a typical
$dE/dx$ distribution normalized by the pion $dE/dx$ at
4.5 $<$ $p_{\mathrm T}$ $<$ 5.0 GeV/$c$ and $\mid\eta\mid$ $<$ 0.5.
The Gaussian distribution used to extract the pion yield and the pion, proton
and anti-proton peak positions are also shown in the figure.

The kaon contamination is estimated via either of the equations given below.
The uncorrected proton yield is
$$p = (Y-\beta (h-\pi))/{(\alpha-\beta)}$$ or
$$p = (Y-\beta K^{\mathrm 0}_{\mathrm S})/{\alpha},$$ where $\alpha$
and $\beta$ are the proton and kaon efficiencies from the
integration described above, derived from the $dE/dx$ calibration,
resolution and the Bichsel function~\cite{rdEdx,bichsel}.  In the
first case the kaon contamination is estimated through the yields of
the inclusive hadrons ($h$) and pions, in case two from known yields
from $K^{\mathrm 0}_{\mathrm S}$ measurements~\cite{rdEdx,MHeinz}.
The typical values of $\alpha$ for a dE/dx cut slightly away from
the proton peak position is 0.4 and the $\beta$ values decrease
from 0.2 to 0.08 with $p_{\mathrm T}$ in the range 2.5 $<$
$p_{\mathrm T}$ $<$ 10 GeV/$c$. At high $p_{\mathrm T}$, the yields
of other stable particles (i.e., electrons and deuterons) are at
least two orders of magnitude lower than those of pions, and are
negligible in our studies.  The two results are consistent where
STAR $K^{\mathrm 0}_{\mathrm S}$ measurements are available.  The
$p_{\mathrm T}$-dependence of the reconstruction efficiency,
background and the systematic uncertainties for pions, protons and
anti-protons for low $p_{\mathrm T}$ in $p$+$p$ and $d$+Au
collisions are described in Ref.~\cite{star_tof}. At high
$p_{\mathrm T}$ ($>$ 2.5 GeV/$c$), the efficiency is almost
independent of $p_{\mathrm T}$ in both $p$+$p$ and $d$+Au
collisions. The tracking efficiencies are $\sim$ 88\% and
92\% in $p$+$p$ and $d$+Au collisions, respectively.
The difference in tracking efficiency arises because of 
worse vertex determination in $p$+$p$ collisions than $d$+Au collisions.
The background contamination to pion spectra for $p_{\mathrm T}$ $>$
2.5 GeV/$c$, primarily from $K^{\mathrm 0}_{\mathrm S}$ weak decay
is estimated from PYTHIA/HIJING simulations with full GEANT detector
descriptions to be $\sim$ 4\%. The charged pion spectra are
corrected for efficiency and background effects. The inclusive
proton and anti-proton spectra are presented with efficiency
corrections and without hyperon feed-down corrections. 
The integrated $\Lambda$/$p$-ratio is estimated to 
be $<$ 25\%~\cite{star_tof,MHeinz}.
Additional corrections are applied for primary vertex reconstruction
inefficiency as discussed in
Refs.~\cite{star_rdau,star_pid,star_tof}. The momentum resolution is
given as $\Delta{p_{\mathrm T}}$/$p_{\mathrm T}$ = 0.01 +
0.005$p_{\mathrm T}$/(GeV/$c$) and has $<$ 4\% effect on the yields
at the highest $p_{\mathrm T}$ value. The spectra are not corrected
for momentum resolution effects, but they are included in the
systematic errors.

%%%%%%%%%%%%%%%%%%%%%%%%%%%%%%%%
%Systematic errors
%%%%%%%%%%%%%%%%%%%%%%%%%%%%%%%%
The total systematic uncertainties associated with pion yields are
estimated to be ${}^<_\sim$ 15\%. This systematic uncertainty is
dominated by the uncertainty in modeling the detector response in
the Monte Carlo simulations. Protons from hyperon ($\Lambda$ and
$\Sigma$) decays away from the primary vertex can be reconstructed
as primordial protons at a slightly higher $p_{\mathrm T}$ than
their true value, but with worse momentum resolution. This results
in an uncertainty of the inclusive proton yield of $\sim2\%$ at
$p_{\mathrm T}$ = 3 GeV/$c$ and $\sim10\%$ at $p_{\mathrm T}$ = 10
GeV/$c$.  For proton and anti-proton yields at high $p_{\mathrm T}$
an additional systematic error arises from the uncertainties in the
determination of the efficiencies, $\alpha$ and $\beta$, under a
specific $dE/dx$ selection for integration. This is due to the
uncertainties in the mean $dE/dx$ positions for protons and kaons.
The total systematic uncertainty in obtaining the proton and
anti-proton yields for $p_{\mathrm T}$ $>$ 2.5 GeV/$c$ increases
with $p_{\mathrm T}$ from 12\% to 23\% (at $p_{\mathrm T}$ = 10
GeV/c) in both $p$+$p$ and $d$+Au collisions. The errors shown in
the figures are statistical, and the systematic errors are
plotted as shaded bands.  In addition, there are overall
normalization uncertainties from trigger and luminosity in $p$+$p$
and $d$+Au collisions of 14\% and 10\%,
respectively~\cite{star_rdau}. These errors are not shown.

%%%%%%%%%%%%%% Fig. 1 %%%%%%%%%%%%%%%%%%%%%%%%%%%%%
\begin{figure}
\begin{center}
\includegraphics*[scale=0.71]{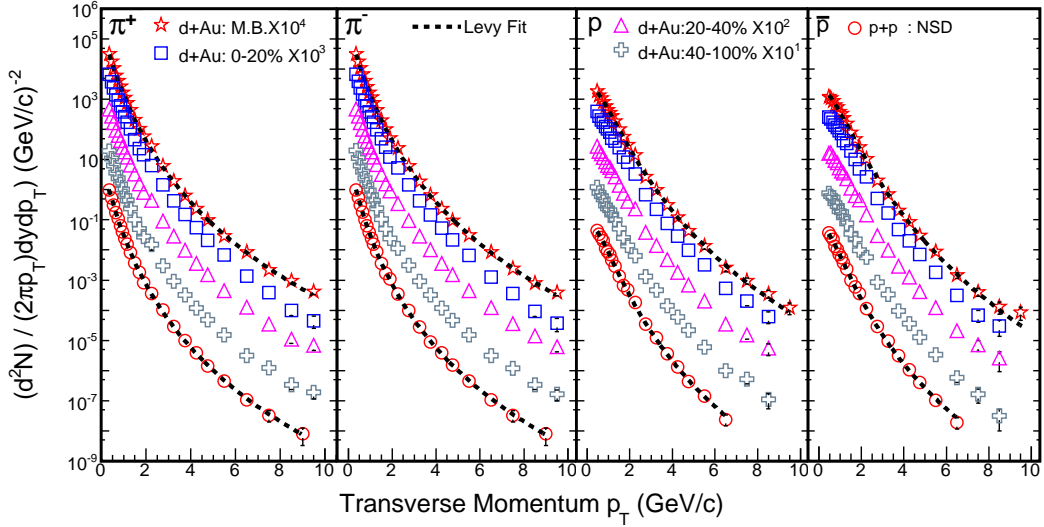}
%\vspace{-1.0 cm}
\caption{Midrapidity ($\mid$$y$$\mid$ $<$ 0.5) transverse momentum
spectra for charged pions, proton and anti-proton in $p$+$p$ and $d$+Au
collisions for various event centrality classes. Minimum bias
distributions are fit to Levy functions which are shown as dashed curves.
}
\label{fig1}
\end{center}
\end{figure}
%%%%%%%%%% End of Fig.1 %%%%%%%%%%%%%%%%%%%%%%%%%%%%%
Figure~\ref{fig1} shows the invariant yields of charged pions,
protons and anti-protons for the $p_{\mathrm T}$ range of 0.3 $<$
$p_{\mathrm T}$ $<$ 10 GeV/$c$ in minimum bias $p$+$p$ collisions
and for various centrality classes in $d$+Au collisions. The yields
span over eight orders of magnitude. The minimum bias distributions
are fit with a Levy distribution~\cite{Levy} of the form
$\frac{d^{2}N}{2\pi p_{T}dp_{T}dy} = \frac{B}{(1 +
(m_{T}-m_{0})/nT)^{n}}$,
where $m_{T}$ = $\sqrt{p_{\mathrm T}^{2}+m_{0}^{2}}$ and $m_{0}$ is the
mass of the hadron.
The Levy distribution essentially takes
a power-law form at higher $p_{\mathrm T}$ and has an exponential
form at low $p_{\mathrm T}$. For the $p$ and $\bar{p}$ spectra, 
fit with a power-law function gives a worse $\chi^{2}/ndf$ 
compared to the fit with the Levy function. For $d$+Au collisions 
the $\chi^{2}/ndf$ for the power-law fit to $p$($\bar{p}$) spectra 
is 68.55/20(86.77/20) and the corresponding value for the fit with 
the Levy function is 21.19/20(26.4/20).

\section{Nuclear modification factor}
The nuclear modification factor ($R_{\mathrm {dAu}}$) can be used to
study the effects of cold nuclear matter on particle production.
It is defined as a ratio of the invariant yields of the produced
particles in $d$+Au collisions to those in $p$+$p$ collisions scaled
by the underlying number of nucleon-nucleon binary collisions.
%\begin{displaymath}
\begin{equation}
%\nonumber
R_{\rm{dAu}}(p_{\rm T})\,=\,\frac{d^2N_{\rm{dAu}}/dy dp_{\rm T}}
{\langle N_{\rm {bin}}\rangle /\sigma_{\rm{pp}}^{\rm {inel \cdot}}\,d^2\sigma_{\rm{pp}}/dy dp_{\rm T}},
%\end{displaymath}
\end{equation}
where $\langle N_{\mathrm {bin}}\rangle$ is the average number
of binary nucleon-nucleon (NN) collisions per event, and
$\langle N_{\mathrm {bin}}\rangle /\sigma_{\rm{pp}}^{\mathrm {inel}}$
is the nuclear overlap function $T_A(b)$~\cite{star_rdau,star_pid}.
The value of $\sigma_{\rm{pp}}^{\rm {inel}}$ is taken to be 42 mb.
%%%%%%%%%%%%%% Fig. 2 %%%%%%%%%%%%%%%%%%%%%%%%%%%%%
\begin{figure}
\begin{center}
\includegraphics[scale=0.48]{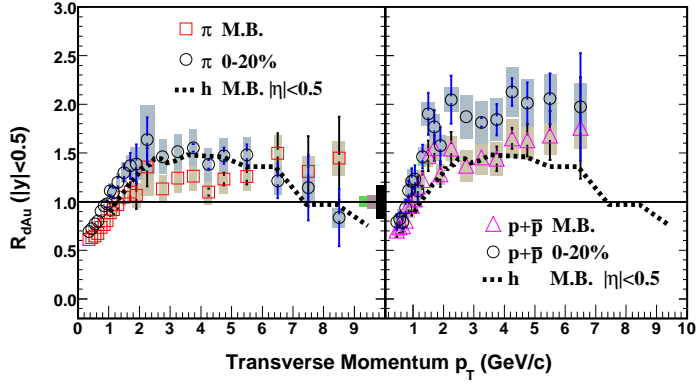}
\caption{
Nuclear modification factors, $R_{\mathrm dAu}$, for charged pions
$\pi^{+}+\pi^{-}$ and $p$+$\bar{p}$ at $\mid$$y$$\mid$ $<$ 0.5
in minimum bias and 0-20\% central $d$+Au collisions.
For comparison results on
inclusive charged hadrons (STAR) from Ref.~\cite{star_rdau}
at $\mid$$\eta$$\mid$ $<$ 0.5 are shown by dashed curves.
The first two shaded bands around $1$ correspond to the error due to
uncertainties in estimating the number of binary collisions in
minimum bias and 0-20\% central $d$+Au collisions respectively.
The last shaded band is the normalization uncertainty from
trigger and luminosity in $p$+$p$ and $d$+Au collisions.
}
\label{fig2}
\end{center}
%\vspace{-13.1pc}
\end{figure}
%%%%%%%%%% End of Fig.2 %%%%%%%%%%%%%%%%%%%%%%%%%%%%%

The left panel of Fig.~\ref{fig2} shows $R_{\mathrm {dAu}}$ values for
charged pions (($\pi^{+}$+$\pi^{-}$)/2) in minimum bias and 0-20\%
central collisions at $\mid$y$\mid$ $<$ 0.5. The $R_{\mathrm {dAu}}$
for 0-20\% central collisions are higher than $R_{\mathrm {dAu}}$ for
minimum bias collisions.  The result $R_{\mathrm {dAu}}$ $>$ 1 indicates
a slight enhancement of high $p_{\mathrm T}$ charged pion yields
in $d$+Au collisions compared to binary collision scaled charged pion
yields in $p$+$p$ collisions within the measured ($y$, $p_{\mathrm T}$)
range.  The right panel of Fig.~\ref{fig2} shows the $R_{\mathrm
{dAu}}$ of baryons ($p$+$\bar{p}$) for the minimum bias collisions at
$\mid$y$\mid$ $<$ 0.5.  The $R_{\mathrm {dAu}}$ for $p+ \bar{p}$ is
again greater than unity for $p_{\mathrm T}$ $>$ 1.0 GeV/$c$ and is
larger than $R_{\mathrm {dAu}}$ for charged pions. The $R_{\mathrm {dAu}}$
of pions for 2$<$ $p_{\mathrm T}$ $<$ 5 GeV/$c$ is 1.24 $\pm$ 0.13 and
that for $p$+$\bar{p}$ is 1.49 $\pm$ 0.17 in minimum bias collisions.
Identified hadron $R_{\mathrm {dAu}}$ are sensitive to nuclear modification
of the PDF from processes such as nuclear shadowing and parton
saturation as well as to transverse momentum broadening, 
energy loss in cold nuclear matter and
hadronization through recombination, thereby further constraining
the models~\cite{vogt}.

\section{Particle ratios}
%%%%%%%%%%%%%% Fig. 3 %%%%%%%%%%%%%%%%%%%%%%%%%%%%%
\begin{figure}
\begin{center}
\includegraphics[scale=0.45]{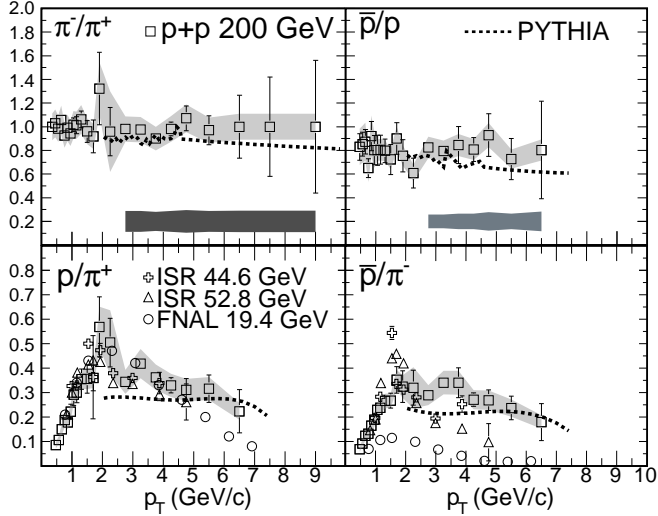}
\caption{
Ratio of $\pi^{-}/\pi^{+}$, $\bar{p}/p$, $p/\pi^{+}$,
$\bar{p}/\pi^{-}$ at midrapidity ($\mid$$y$$\mid$ $<$ 0.5) as a
function of $p_{\mathrm T}$ in $p$+$p$
minimum bias collisions.
For comparison the results from lower energies at ISR~\cite{isr53}
and FNAL~\cite{fnal} are also shown for $p/\pi^{+}$ and
$\bar{p}/\pi^{-}$ ratios. The dotted curves are the results from PYTHIA.
The shaded bands below the $\pi^{-}/\pi^{+}$ and $\bar{p}/p$ ratios are the point--to--point
correlated errors in the yields associated with the ratio.
}
\label{fig3a}
\end{center}
\end{figure}
%%%%%%%%%% End of Fig.3 %%%%%%%%%%%%%%%%%%%%%%%%%%%%%

\begin{figure}
\begin{center}
\includegraphics[scale=0.45]{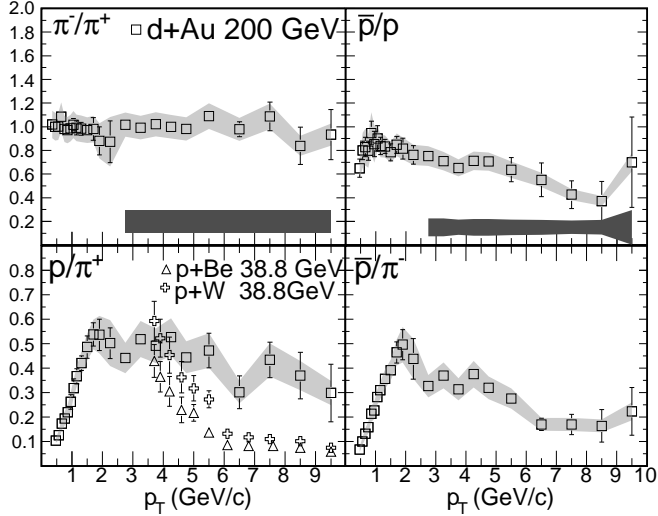}
\caption{ Same as Fig.~\ref{fig3a} for d+Au minimum bias collisions.
For comparison the $p/\pi^{+}$-ratio  from lower energies at
FNAL~\cite{fnal} are shown.}
\label{fig3b}
\end{center}
\end{figure}

The particle ratios at midrapidity as a function of $p_{\mathrm T}$
for $p$+$p$ and $d$+Au minimum bias collisions are shown in
Figs.~\ref{fig3a} and ~\ref{fig3b} respectively. Correlated
errors are shown as the shaded bands below the data points. The
$\pi^{-}/\pi^{+}$-ratio has a value $\sim$ 1 and is independent of
$p_{\mathrm T}$ in both $p$+$p$ and $d$+Au collisions. The
$\bar{p}/p$-ratio for $p$+$p$ collisions is also independent of
$p_{\mathrm T}$ within the range studied and has a value of 0.81
$\pm$ 0.1 at 2.5 $<$ $p_{\mathrm T}$ $<$ 6.5 GeV/$c$. However, in
$d$+Au collisions we observe a clear decrease of $\bar{p}/p$ for
$p_{\mathrm T}>$6 GeV/$c$.  In quark fragmentation, the leading
hadron is more likely to be a particle rather than an anti-particle,
and there is no such preference from a gluon jet.  A decrease in the
antiparticle/particle ratio with $p_{\mathrm T}$ would then indicate
a significant quark jet contribution to the baryon production.  It
is, however, not clear whether the same effect exists in $p$+$p$
collisions or whether the decrease of $\bar{p}/p$ is due to
additional nuclear effects in $d$+Au collisions.
Calculations from PYTHIA(v6.319) predict
somewhat more prominent $p_{\mathrm T}$-dependence~\cite{pythia}.

%%%%%%%%%%%%%% Fig. 3 %%%%%%%%%%%%%%%%%%%%%%%%%%%%%
%%%%%%%%%% End of Fig.3 %%%%%%%%%%%%%%%%%%%%%%%%%%%%%

At RHIC, the $p/\pi^{+}$ and $\bar{p}/\pi^{-}$
ratios increase with $p_{\mathrm T}$ up to 2 GeV/$c$ and then start
to decrease for higher $p_{\mathrm T}$ in both $p$+$p$ and $d$+Au
collisions.
The $\bar{p}/\pi^{-}$-ratio rapidly approaches a value of
0.2, which is  between the values in $e^{+}$+$e^{-}$ collisions for
quark and gluon jets~\cite{delphi,opal}.
The $p/\pi^{+}$ and $\bar{p}/\pi^{-}$ ratios from PYTHIA
are constant at high $p_{\mathrm T}$ in contrast to a
decreasing trend oberserved in the data.
The $p/\pi^{+}$-ratios in $p$+$p$ collisions compare
well with results from lower energy ISR and FNAL fixed target
experiments~\cite{isr53,fnal}. Meanwhile, $\bar{p}/\pi^{-}$-ratios at high
$p_{\mathrm T}$ have a strong energy dependence with larger values at
higher beam energies. In $d$+Au collisions the $p/\pi^{+}$-ratio at
high $p_{\mathrm T}$ is lower for $p$+A collisions at FNAL energy than at RHIC.

\section{Comparison to NLO pQCD and model calculations}
%%%%%%%%%%%%%% Fig. 4 %%%%%%%%%%%%%%%%%%%%%%%%%%%%%
\begin{figure}
\begin{center}
\includegraphics*[scale=0.34]{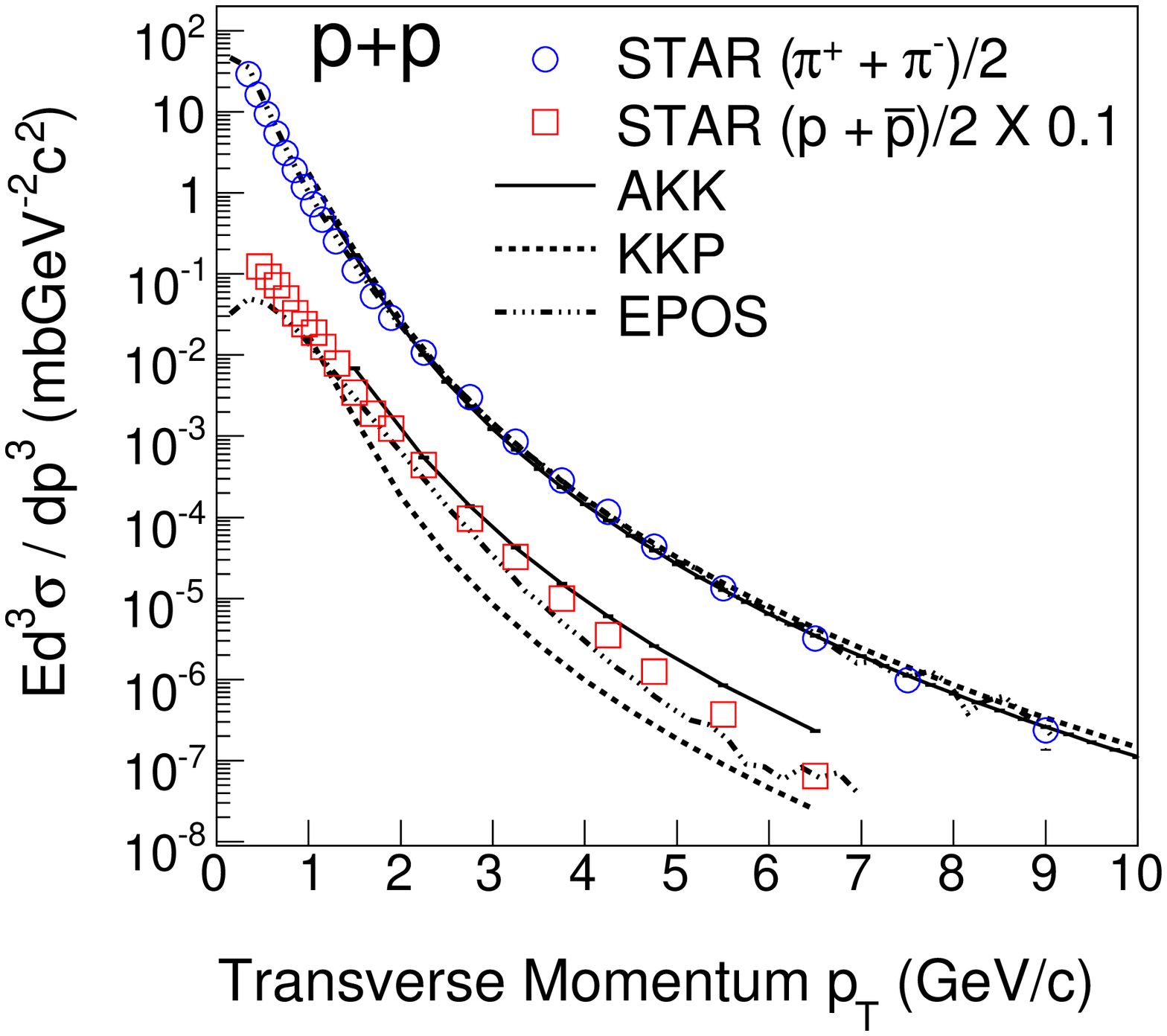}
\includegraphics*[scale=0.34]{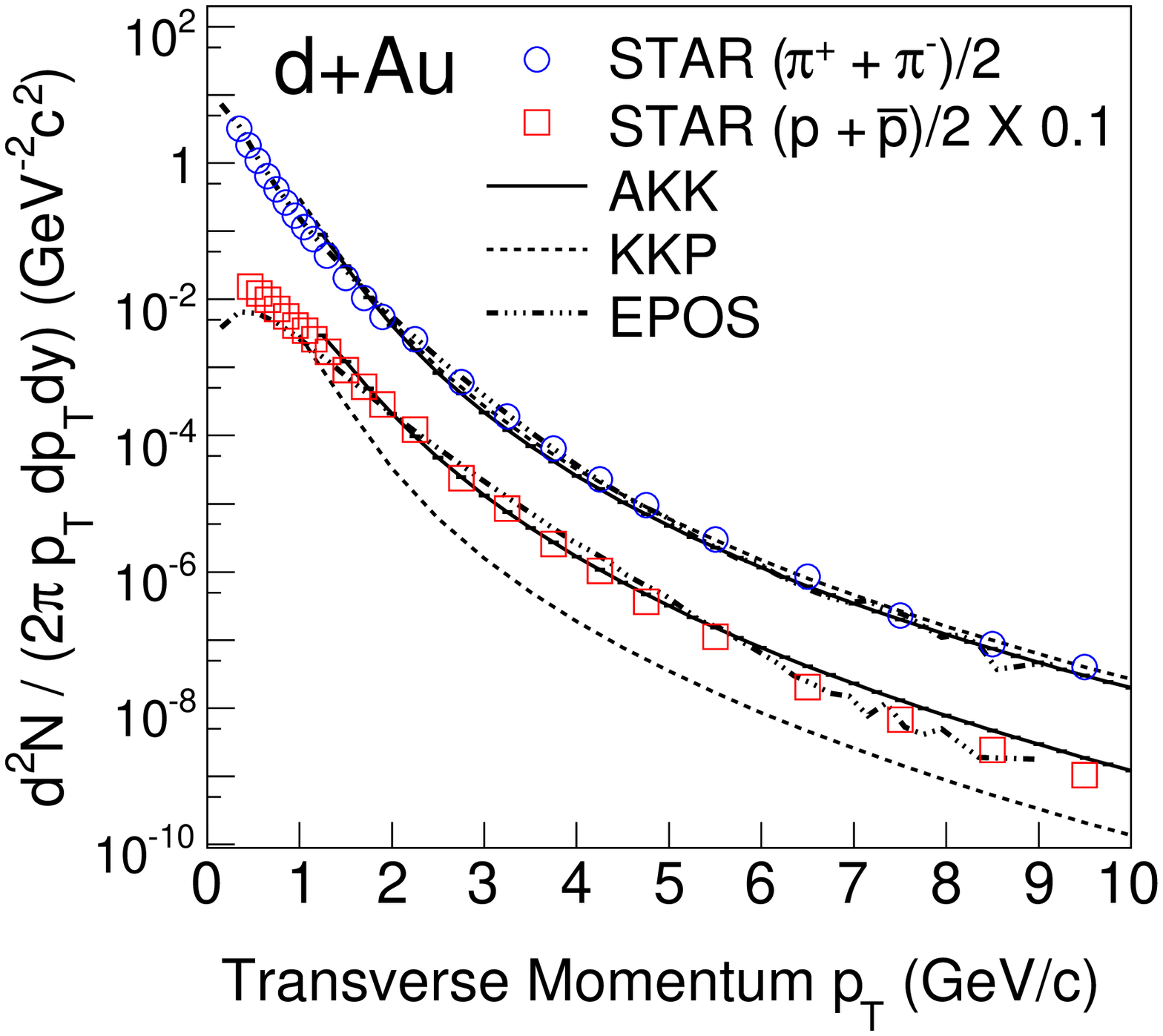}
\caption{Midrapidity invariant yields for ($\pi^{+}$ +
$\pi^{-}$)/2 and ($p$+$\bar{p}$)/2 at high $p_{\mathrm T}$ for
minimum bias $p$+$p$ and $d$+Au collisions compared to results from
NLO pQCD calculations using
KKP~\cite{pqcd_kkp} (PDF: CTEQ6.0) and AKK~\cite{pqcd_akk} (PDF:
CTEQ6M) sets of fragmentation functions and results from the EPOS
model~\cite{epos}. The PDFs for $d$ and Au-nucleus are taken
from Refs.~\cite{pdf_p} and \cite{pqcd_dau} respectively.
All results from NLO pQCD 
calculations are with factorization scale is $\mu$ = $p_{\mathrm T}$.
}
\label{fig4}
\end{center}
\end{figure}
%%%%%%%%%% End of Fig.4 %%%%%%%%%%%%%%%%%%%%%%%%%%%%%

In Fig.~\ref{fig4} we compare ($\pi^{+}$ + $\pi^{-}$)/2 
and ($p$+$\bar{p}$)/2 yields in minimum bias $p$+$p$
and $d$+Au collisions at midrapidity for high $p_{\mathrm T}$ to those from
NLO pQCD calculations and the phenomenological parton model 
(EPOS)~\cite{epos}. The results from EPOS agree fairly well
with our data for
charged pions and proton+anti-proton in $p$+$p$ and $d$+Au
collisions.  The NLO pQCD results are based on calculations
performed with two sets of FFs, the {\it Kniehl-Kramer-Potter
(KKP)}~\cite{pqcd_kkp} and the {\it Albino-Kniehl-Kramer (AKK)} set
of functions~\cite{pqcd_akk}. The factorization scale for all the
NLO pQCD calculations shown is for $\mu$ = $p_{\mathrm T}$. The
charged pion data for $p_{\mathrm T}$ $>$ 2 GeV/$c$ in $p$+$p$
collisions are reasonably well described by the NLO pQCD
calculations using the KKP and AKK set of FFs. 
A similar observation for $\pi^{0}$s using KKP FFs
was made by the PHENIX Collaboration~\cite{phenix_pp}. For $d$+Au
collisions NLO pQCD calculations with KKP FFs are consistent with
the data for $p_{\mathrm T}$ $>$ 4 GeV/$c$ while those with AKK FFs
underpredict the measured charged pion yields.

The proton+anti-proton yield at high $p_{\mathrm T}$
in $p$+$p$ and $d$+Au collisions is much higher than the results from
NLO pQCD calculations using the KKP set of FFs and lower compared to
calculations using AKK FFs.
The relatively better agreement of NLO pQCD calculations
with AKK FFs compared to those with KKP FFs for proton+anti-proton
yields shows the importance of the flavor-specific measurements in
$e^{+}$+$e^{-}$ collisions in determining the FFs for baryons. One may
further improve the NLO pQCD calculations by an all-order resummation
of large logarithmic corrections to the partonic
cross-sections~\cite{vogalsen}.

\section{Scaling of particle production}
The invariant cross-sections of inclusive pion production in high
energy $p$+$p$ collisions have been found to follow the scaling laws~\cite{scaling} :
\begin{equation}
E \frac{d^3\sigma}{dp^3}=\frac{1}{p_T^{n}} f\left({x_T}\right) \quad \mbox{or}\qquad
E \frac{d^3\sigma}{dp^3}=\frac{1}{\sqrt{s}^{n}} g\left({x_T}\right)
\label{eq:bbg}
\end{equation}
where $x_{\mathrm T}$ = 2$p_{\mathrm T}/\sqrt{s}$ and $f\left({x_T}\right)$ and $g\left({x_T}\right)$ 
are some functions of $x_{\mathrm T}$. Similar scaling has
been observed in $e^{+}$+$e^{-}$ collisions, but without the
$\sqrt{s}^{n}$ or $p_{T}^{n}$ factor~\cite{escaling}.  The value of
the power $n$ ranges from 4 to 8~\cite{scaling_theory}. In the general
scaling form $\sim$ 1/$p_{\mathrm T}^{n}$, $n$ depends on the quantum
exchanged in the hard scattering. In parton models, it is related to
the number of point-like constituents taking an active role in the
interaction. The  value reaches 8 in the case of a quark-meson
scattering by exchanging a quark.  With the inclusion of QCD, the
scaling law follows as $\sim$ 1/$\sqrt{s}^{n}$, where $n$ becomes a
function of $x_{\mathrm T}$ and $\sqrt{s}$.  The value of $n$ depends
on the evolution of the structure function and FFs.
%%%%%%%%%%%%%% Fig. 5 %%%%%%%%%%%%%%%%%%%%%%%%%%%%%
\begin{figure}
\begin{center}
\includegraphics[scale=0.75]{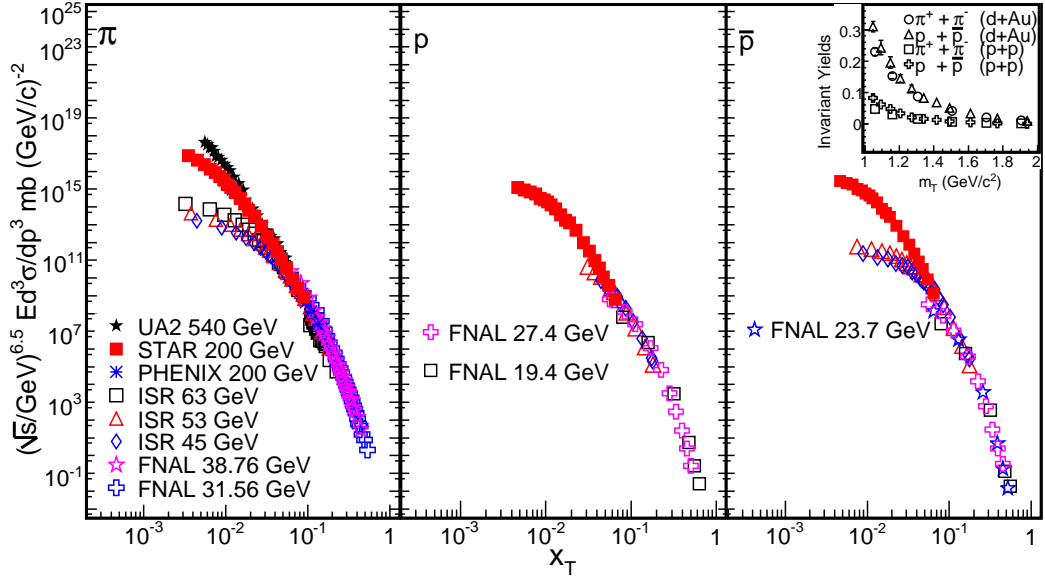}
\caption{$x_{T}$-scaling of pions, protons and anti-protons. The
data from other experiments are from the following references, FNAL
: Refs~\cite{fnal,cronin}, ISR : Ref.~\cite{isr53}, PHENIX :
Ref.~\cite{phenix_pp}, and UA2~\cite{ua2}. The inset shows the
$m_{\mathrm T}$-scaling of the invariant yields for charged pions
and protons+anti-protons in $p$+$p$ and $d$+Au collisions.
} \label{fig5}
\end{center}
\end{figure}
%%%%%%%%%% End of Fig.5 %%%%%%%%%%%%%%%%%%%%%%%%%%%%%
$n$=4 is expected in more basic scattering processes
(as in QED) ~\cite{scaling_theory,diquark}.

Figure~\ref{fig5} shows
the $x_{\mathrm T}$-scaling of pions, protons and anti-protons.  The
value of $n$ obtained for the scaling with $\sqrt{s}^{n}$ of
the invariant cross-section is 6.5 $\pm$ 0.8. The STAR data covers
the range 0.003 $<$ $x_{\mathrm T}$ $<$ 0.1. The data points
deviate from the scaling behavior for $p_{\mathrm T}<2$ GeV/$c$ for pions and protons, 
which could be interpreted as a transition region from soft to hard
processes in the particle production. The deviations start at a higher
$p_{\mathrm T}$ for the anti-protons.  The available data on pion and
proton invariant cross-sections at various center-of-mass
energies~\cite{isr53,fnal,phenix_pp,cronin,scaling,ua2} for
$p_{\mathrm T}$ $>$ 2 GeV/$c$ are compiled and fitted using the
function $\frac{1}{p_{\mathrm T}^{n}} \left({1 - x_{\mathrm
T}}\right)^{m}$.  The value of $n$ ranges from 6.0 to 7.3 for
$\sqrt{s_{\mathrm {NN}}}$ between 19 GeV and 540 GeV, while that for
$m$ ranges between 13 and 22.  The average value of $n$ for pions is
6.8 $\pm$ 0.5 and that for protons and anti-protons is 6.5 $\pm$
1.0. The variations in $n$ and $m$ values may lead to differences in
details of scaling behaviour at different energies when the cross-section
is multiplied by $1/p_{\mathrm T}^{n}$~\cite{brodsky}. This feature is not
observed in the scaling shown in Fig.~\ref{fig5} due to the data spanning
several orders of magnitude.  The inset of Fig.~\ref{fig5} shows
the $m_{\mathrm T}$ scaling at $p_{\mathrm T}<2$ GeV/$c$,
consistent with possible transition between
soft and hard processes at around $p_{\mathrm T}\simeq2$ GeV/$c$.
The $m_{\mathrm T}$-scaling also indicates that flow 
effects in $p$+$p$ and $d$+Au collisions are
negligible~\cite{rhicwhitepapers,hydro}.  The presented data suggests
that the transition region from soft to hard physics occurs
around $p_{\mathrm T}$${\sim}$2 GeV/$c$ in $p$+$p$ collisions.

\section{Summary}
We have presented transverse momentum spectra for identified
charged pions, protons and anti-protons from $p$+$p$ and $d$+Au collisions
at $\sqrt{s_{\mathrm {NN}}}$ = 200 GeV. The transverse momentum
spectra are measured around midrapidity ($\mid$$y$$\mid$ $<$ 0.5)
over the range of 0.3 $<$ $p_{\mathrm T}$ $<$ 10 GeV/$c$ with particle
identification from the ionization energy loss and its relativistic
rise in the Time Projection Chamber, as well as the Time-of-Flight in STAR.
The following conclusions can be drawn from the
present study:
(a) The nuclear modification factor around midrapidity is
     enhanced in $d$+Au collisions to about 1.5 for pions and to about
     2 for protons and antiprotons at intermediate 
    $p_{\mathrm T}$ (2 $<$ $p_{\mathrm T}$ $<$ 5 GeV/$c$). 
%   The identified particle nuclear modification
%    factor around midrapidity is enhanced
%    above unity for pions, and the effect on proton
%    and anti-proton spectra is even larger at the
%    intermediate $p_{\mathrm T}$ (2 $<$ $p_{\mathrm T}$ $<$ 5 GeV/$c$).
(b) Identified particle ratios were measured up to $p_{\mathrm T}$ of 7 GeV/$c$
     in $p$+$p$ and 10 GeV/$c$ in $d$+Au reactions. Their dependence on 
     species, $p_{\mathrm T}$ and collisions energy was shown to be 
     sensitive to the relative contributions from quark and gluon 
     fragmentation as well as to their fragmentation functions.
%    Particle ratios as function of beam energy, beam species and
%   transverse momentum reveal their dependence on the FFs and the
%   relative quark and gluon contributions.
(c) The NLO pQCD calculations describe the high $p_{\mathrm T}$ data for
    charged pions reasonably well in $p$+$p$ collisions and $d$+Au
    collisions. In general, baryon production has historically been difficult
    to describe by pQCD and hadronization~\cite{diquark,hhp}.
    Use of the recently published AKK FFs results in a much improved 
    description of the measured $p$ and $\bar{p}$ spectra.
%    The improved description of experimental data in RHIC's $p$+$p$ collisions
%    by the AKK FFs, which come from NLO pQCD fits to the $e^{+}$+$e^{-}$
%    data, is of great importance.
(d) The proton and pion spectra in $p$+$p$ collisions
    follow $x_{\mathrm T}$-scaling with a beam-energy dependent factor
    $\sim$ $\sqrt{s_{\mathrm {NN}}}^{6.5}$ above
    $p_{\mathrm T}$$\sim$ 2 GeV/$c$. The pion and proton spectra
    follow transverse mass scaling for
     $m_{\mathrm T}$ $<$ 2 GeV/$c^{2}$ in both $p$+$p$ and
    $d$+Au collisions, suggesting the transition region from soft
    to hard process domination occurs at $p_{\mathrm T}$$\sim$ 2 GeV/$c$ in
    these collision systems.
     The measurements presented in this paper provide better
     constraints on jet quenching and quark recombination models which
     are presently the best candidates for explaining particle
     production in the intermediate $p_{\mathrm T}$ region.
%   These findings will provide a better foundation for
%    constraining applications of jet quenching and quark
%    recombination models to explain the phenomena in A+A collisions
%    in this $p_{\mathrm T}$ range.

We would like to thank Simon Albino, Stefan Kretzer and Werner Vogelsang
for providing us the NLO pQCD results, Klaus Werner for the
EPOS results and J. Raufeisen for useful discussions. We thank the
RHIC Operations Group and RCF at BNL, and the NERSC Center at LBNL
for their support. This work was supported in part by the HENP
Divisions of the Office of Science of the U.S. DOE; the U.S. NSF;
the BMBF of Germany; IN2P3, RA, RPL, and EMN of France; EPSRC of the
United Kingdom; FAPESP of Brazil; the Russian Ministry of Science
and Technology; the Ministry of Education and the NNSFC of China;
IRP and GA of the Czech Republic, FOM of the Netherlands, DAE, DST,
and CSIR of the Government of India; Swiss NSF; the Polish State
Committee for Scientific Research; STAA of Slovakia, and the Korea
Sci. \& Eng. Foundation.

\end{document}